\documentclass[11pt,a4paper,twocolumn]{article}
\usepackage{qtree,amstext,amsthm,graphics,url}

\title{\textbf{Towards practical meta-querying}}

\author{\begin{tabular}{cc} \textbf{Jan Van den Bussche}$^{1}$ &
    \textbf{Stijn Vansummeren}$^{1}$ \\
    \multicolumn{2}{c}{\textbf{Gottfried Vossen}$^{2}$}\\[3pt]
    $^{1}$University of Limburg, Belgium & $^{2}$University of Muenster,
    Germany
\end{tabular}}

\date{}

\begin{document}

\maketitle

\begin{abstract}

% \begin{center}
% \textbf{Abstract}

% \bigskip
% \parbox{75ex}{
We describe a meta-querying system for databases containing queries in
addition to ordinary data.  In the context of such databases, a
meta-query is a query about queries.  Representing stored queries in
XML, and using the standard XML manipulation language XSLT as a
sublanguage, we show that just a few features need to be added to SQL
to turn it into a fully-fledged meta-query language.  The good news
is that these features can be directly supported by extensible
database technology.
% }
% \end{center}
\end{abstract}

\section{Introduction}

Enterprise databases often contain not only ordinary data, but also
queries. Examples are view definitions in the system catalog; usage logs
or workloads; and stored procedures as in SQL/PSM or Postgres
\cite{stoneb_procedures}.
Unfortunately, these queries are typically stored as long strings,
which makes it hard to use standard SQL to express \emph{meta-queries:}
queries about queries.  Meta-querying is an important activity in situations
such as advanced database administration, database usage monitoring, and
workload analysis.
Examples of meta-queries to a usage log are:
\begin{enumerate}
\item
Which queries in the log do the most joins?
\item
Which queries in the log return an empty answer on the current instance of the
database?
\item
View expansion: replace, in each query in the log,
each view name by its definition as given in the system catalog.
\item
Given a list of new view definitions (under the old names),
which queries in the log give a different answer on the
current instance under the new view definitions?
\end{enumerate}
Query~1 is \emph{syntactical:}
it only queries the stored queries on the basis of
their expressions.  Query~2 is \emph{semantical:}
its answer depends on the results
of dynamically executing the stored queries.
Query~3 is again syntactical, but more so than
query~1 in that it also performs syntactical transformations.
Query~4 is syntactical and semantical together.

To express meta-queries, database administrators and other advanced
users typically resort to a programming language like Perl, in
combination with Dynamic SQL\@.  It would be much nicer if they would
not have to ``leave'' the database system and could express their
meta-queries directly in Interactive SQL\@.  Indeed, already in 1993,
in his SIGMOD Innovations Award speech, Jim Gray urged the database
community to lower the wall between data and programs.  In the same vein, the
Asilomar Report puts the unification between programs and data high on
the database research agenda \cite{asilomar}.  As queries are an
important kind of program in the context of databases, support for
meta-querying thus seems to be a step in the right direction towards
understanding how we can unify programs and data in database systems.

In this paper, we present a practical meta-querying system based on
the relational model. Our main design goal was
to use current DBMS technology and
only extend standard SQL with specific meta-querying features where
necessary. Stored queries are represented as syntax trees in XML
format.  This representation provides a convenient basis for
syntactical meta-querying. Indeed, rather than reinventing the wheel
and designing a new sublanguage for syntactical manipulation of stored
queries, it allows us to use the standard XML transformation language
XSLT for this purpose.  Many syntactical meta-queries can then
directly be expressed simply by allowing XSLT function calls within
SQL expressions.\footnote{We embrace XSLT because it is the most
  popular and stable standard general-purpose XML manipulation
  language to date.  When other languages, notably XQuery
  \cite{xquery}, will take over this role, it will be an easy matter
  to substitute XSLT by XQuery in our overall approach.}

This combination of SQL and XSLT gives us a basic level of expressive
power, but for more complex syntactical meta-queries we need a bit more.
To this end, we enrich the SQL language with \emph{XML variables} which
come in addition to SQL's standard range variables. XML variables range
not over the rows of a table, but rather over the subelements of an XML
tree. The range can be narrowed by an XPath expression. (XPath is the
sublanguage of XSLT used to locate subelements of XML documents.)
XML variables thus
allow us to go from an XML document to a set of XML documents.
Conversely, we also add \emph{XML aggregation} \cite{almaden_relationalxml},
which allows us to go from a set of XML documents to a single one.

SQL combined with XSLT and enriched with XML variables and aggregation
offers all the
expressive power one needs for ad-hoc syntactical meta-querying.
To allow for semantical meta-querying as well, it now suffices to add an
\emph{evaluation
function}, taking the syntax tree of some query as input, and
producing the table resulting from executing the query as output.
We note that a similar evaluation feature was already present in the
Postgres system.

What we obtain is \emph{Meta-SQL}: a practical meta-query language.
Meta-SQL has as advantage that it is \emph{not}
``yet another query language'': it is entirely
compatible with modern SQL implementations offered by contemporary extensible
database systems.  Indeed, these systems already support calls to
external functions from within SQL expressions, which allows us
to implement the XSLT calls.
Furthermore, XML variables and the evaluation
function can be implemented using \emph{set-valued}
external functions.  As we will show,
the powerful feature of ``lateral derived tables'',
part of the SQL:1999 standard,
turns out to be crucial to make this work.
XML aggregation, finally, can be implemented as a user-defined aggregate
function.

We emphasize again that we are \textit{not} proposing yet another database
language.  Instead, our main design goal was to stick as closely as possible
to standard SQL\@. Of course, a drastic alternative is to abandon the
relational model altogether and move to, e.g., an XML-XQuery environment,
where meta-querying does not pose any problem.  However, given the widespread
use of relational databases, a conservative approach such as ours remains
important.

This paper is further organized as follows.  In Section~2, we
combine SQL with XSLT\@.  In Section~3, we add XML variables.  In
Section~4, we move on to semantical meta-querying.  In Section~5,
we describe how Meta-SQL can be implemented using extensible
database technology. We give some experimental performance results of
our prototype in Section~6. In Section~7, we conclude with a discussion
of our approach.

\section{SQL + XSLT}

Consider a standard relational database, except that in a table some
columns can be marked to be of type ``XML''\@.  In any row of that table, the
component corresponding to a column of type XML holds an XML
document.
At the present conceptual level, we do not yet care
about how this is implemented.

XSLT \cite{xslt} is a widely used manipulation language for XML documents.
An XSLT program
takes an XML document as input, and
produces as output another XML document (which could be in degenerate form,
holding just a scalar value like a number or a string).
Using the XSLT top-level
parameter binding mechanism \cite{xslt}, programs can also take additional
parameters as input.\footnote{In
this paper, we cannot include a tutorial on XSLT, for which we refer to the
Web or to the literature \cite{xslt_oreilly,xslt_kay}.}

Hence, to query databases containing XML, it seems natural to extend SQL
by allowing calls to XSLT functions, in the same way as extensible database
systems extend SQL with calls to external functions.  However, in these
systems, external functions have to be precompiled and registered before they
can be used.  In Meta-SQL,
the programmer merely includes the source of the needed XSLT
functions and can then call them directly.

\begin{figure*}
\centering
\scalebox{0.75}{\resizebox{\textwidth}{!}{\includegraphics{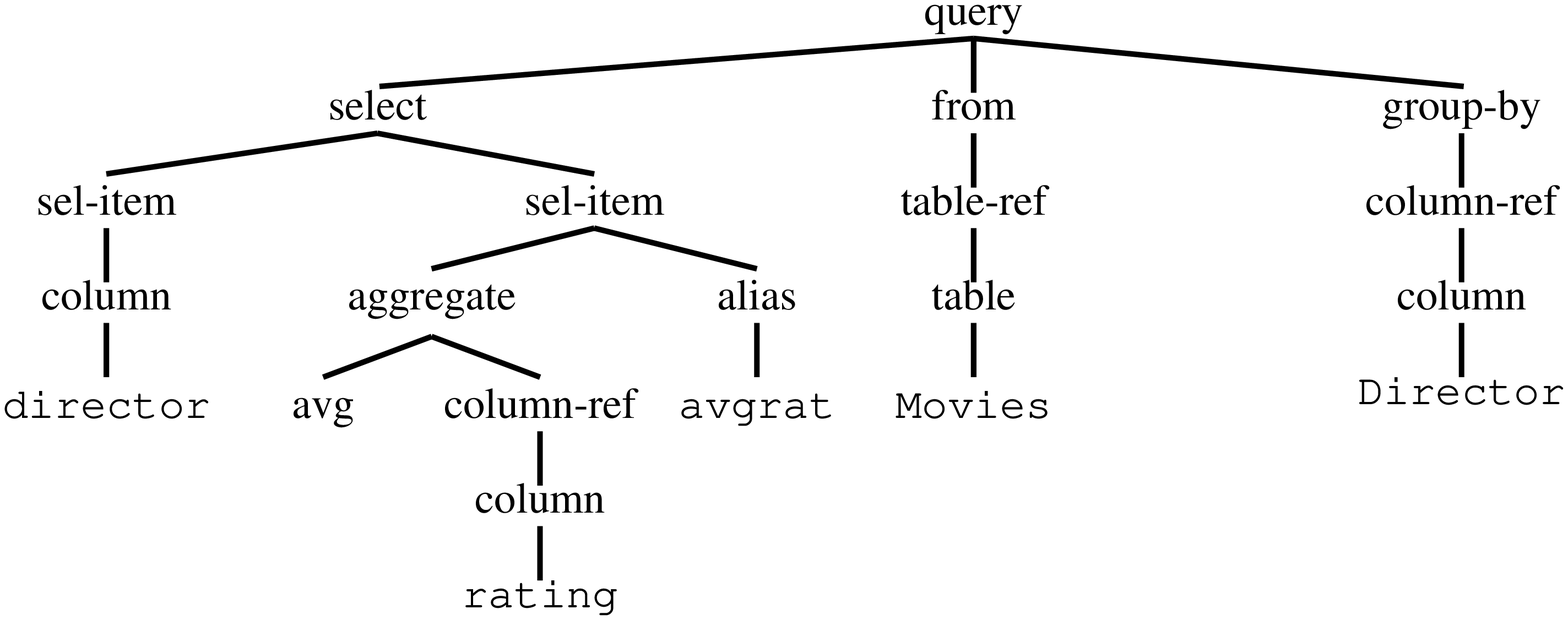}}}
\caption{Syntax tree of SQL query \texttt{select director, avg(rating) as
avgrat from Movies group by director}.}
\label{xmltree}
\end{figure*}

Let us see an example of all this, at the same time applying it
to meta-querying.  Consider a simplified system catalog
table, called \texttt{Views}, containing view definitions.  There is a
column \texttt{name} of type string, holding the view name, and a
column \texttt{def} of type XML, holding the syntax tree of the SQL
query defining the view, in XML format.  For example, over a movies
database, suppose we have a view \texttt{DirRatings} defined as follows:
\begin{verbatim}
select director, avg(rating) as avgrat
from Movies group by director
\end{verbatim}
Then table \texttt{Views} would have a row with value for
\texttt{name} equal to `\texttt{DirRatings}', and value for
\texttt{def} equal to the following XML document:
\begin{verbatim}
<query>
<select>
<sel-item>
  <column>director</column>
</sel-item>
<sel-item>
  <aggregate><avg/>
    <column-ref>
      <column>rating</column>
    </column-ref>
  </aggregate>
  <alias>avgrat</alias>
</sel-item>
</select>
<from>
<table-ref>
  <table>Movies</table>
</table-ref>
</from>
<group-by>
<column-ref>
  <column>director</column>
</column-ref>
</group-by>
\end{verbatim}
Figure~\ref{xmltree} shows the same document as a DOM tree \cite{xmldom},
which is perhaps clearer.

To achieve uniformity in the specific XML format for representing SQL
syntax trees, we must agree on some fixed BNF syntax for SQL\@.
A BNF grammar can be easily transformed into an XML DTD \cite{xml},
which then specifies the XML format to use.
In this paper, we use the BNF grammar given by Date
\cite{date_sql92}.  The derived DTD is given in Appendix~\ref{appdtd}.

Now recall the first example meta-query from the Introduction, but applied to
our \texttt{Views} table:
``which queries do the most joins?''  For simplicity of exposition,
let us identify the
number of joins an SQL query does with the number of table names
occurring in it.  To express this meta-query in Meta-SQL,
we write an auxiliary XSLT function \texttt{count\_tables}, followed by
an obvious SQL query calling this function:
\begin{verbatim}
function count_tables returns number
begin
<xsl:template match="/">
  <xsl:value-of
    select="count(//table)"/>
</xsl:template>
end

select name from Views
where count_tables(def) =
(select max(count_tables(def))
 from Views)
\end{verbatim}
The first line declares the XSLT function in Meta-SQL; between
\texttt{begin} and \texttt{end} one writes arbitrary XSLT code.
Of course, in general, Meta-SQL allows multiple XSLT functions to be
declared and called in the SQL query that follows the function declarations.

As a second example, suppose we are given a list \texttt{Removed}
of names of tables that are going to be removed, and we want to know which
views will become invalid after this removal because they mention one of
these table names.  To express this meta-query in Meta-SQL, we write:
\begin{verbatim}
function mentions_table
param tname string
returns string
begin
<xsl:param name="tname"/>
<xsl:template match="/">
<xsl:if
  test="//table[string(.)=$tname]">
true
</xsl:if>
</xsl:template>
end

select name from Views, Removed
where mentions_table(def,Removed.name)
      = 'true'
\end{verbatim}
Note that, using the XSLT parameter mechanism already mentioned near the
beginning of this section, function \texttt{mentions\_table} takes an extra
parameter \texttt{tname} (of type string). Accordingly,
in the SQL query, the function is called on two arguments: the XML input
\texttt{def}, and the parameter \texttt{Removed.name}.

As a final example, suppose we are given a second view definitions
table \texttt{Views2},
and for every view name that is listed in both views tables, we want
a new definition that equals the union of the first definition and the second
definition.  To express this meta-query in Meta-SQL, we write:
\begin{verbatim}
function unite
param v2 xml
returns xml
begin
<xsl:param name="v2"/>
<xsl:template match="/">
<union>
  <xsl:apply-templates/>
  <xsl:apply-templates select="$v2"/>
</union>
</xsl:template>
<xsl:template match="*">
<xsl:copy>
  <xsl:apply-templates/>
</xsl:copy>
</xsl:template>
end

select name, unite(v.def,v2.def)
from Views v, Views2 v2
where v.name=v2.name
\end{verbatim}
Note that function \texttt{unite} outputs XML, and
again takes an extra parameter \texttt{v2}, now also of type XML.

\section{XML variables and XML aggregation} \label{xmlvaragg}

The simple combination SQL + XSLT is already quite useful, but its
full potential is only realized when we add a language construct that allows
us to extract the subelements of an XML document.  For example, the
simple meta-query
\begin{quote}
give all pairs $(v,t)$, where $v$ is a view name and $t$ is a
table name mentioned in the definition of view $v$
\end{quote}
is otherwise not expressible.

We therefore add \emph{XML variables} to SQL: while the standard SQL
range variables range over the rows of a table, XML variables range over
the subelements of an XML document.  Like range variables, XML variables
are bound in the from-clause, in a similar way variables are bound in
OQL \cite{cluet_oql} and in XQuery \cite{xquery}.
More specifically, an XML variable $x$ is
bound in a from-clause using a construct of the following form:
$$ \texttt{$x$ in $y$[$e$]} $$
Here,
\begin{itemize}
\item
$y$ is a previously bound XML variable or a column reference,
or an XSLT function call, of type XML; and
\item
$e$ is an XPath expression
\cite{xpath} specifying which subelements of $y$ we want $x$ to range
over.
\end{itemize}
A bound XML variable can appear in an SQL expression anywhere a column
reference can.

For example, the meta-query quoted above can now be expressed as follows:
\begin{verbatim}
function string_value returns string
begin
end

select v.name, string_value(x)
from Views v, x in v.def[//table]
\end{verbatim}
Note that the body of function \texttt{string\_value} is empty; indeed, the
empty XSLT program does exactly what we want here, namely, to return the
string value of an XML document (in this case, a \texttt{table} subelement).

As another example, suppose we are given a log table \texttt{Log} with
stored queries (in a column \texttt{Q}), and we want to identify ``hot
spots'': subqueries that occur as a subquery in at least ten different
queries.  To express this meta-query, we write:
\begin{verbatim}
select s
from Log l, s in l.Q[//query]
group by s
having count(l.Q) >= 10
\end{verbatim}

\paragraph{XML aggregation} XML variables allow us to go from an XML document
to a set of XML documents.  Conversely, we want to be able to go from a set
of XML documents to a single one.  Thereto, we add a natural aggregate function
on XML columns, called CMB (for ``combine''), also used by
Shanmugasundaram et al \cite{almaden_relationalxml}.
Just like the result of the
standard aggregate function SUM on a list of numbers $x_1,\dots,x_n$ is the
sum $x_1+\cdots+x_n$, the result of CMB on a list of XML documents
$d_1,\dots,d_n$ is the combined XML document $$ \Tree [.cmb $d_1$ {\dots}
$d_n$ ]
$$  So, the top-level element of ${\rm CMB}(d_1,\dots,d_n)$ is always labeled
`cmb', and has the documents $d_1,\dots,d_n$ as its
subelements.\footnote{CMB is not commutative, so in the
outcome of an XML aggregation, the particular order of the grouped documents
is undetermined and will be implementation-dependent. Shanmugasundaram et al
\cite{almaden_relationalxml} consider an ordered version of CMB.}

As an example, suppose we are given a view definitions table \texttt{Views3}
similar to \texttt{Views}, except that the same view name may appear with
multiple definitions.  Suppose
we want to ask the following meta-query:
for each view name, give the Cartesian product of its
definitions.  Thereto, we first write an easy XSLT function \texttt{cartprod}
(see Appendix~\ref{appxslt})
that transforms an input
of the form ${\rm CMB}(d_1,\dots,d_n)$ into the document $$ \Tree [.query
[.select wildcard ] [.from [.table-ref $d_1$ ] {\dots} [.table-ref $d_n$ ]]] $$
We then write:
\begin{verbatim}
select name, cartprod(CMB(def))
from Views3 group by name
\end{verbatim}

As another example, recall the third example meta-query from the Introduction:
view expansion in the log.  To express this meta-query in Meta-SQL, we first
write two auxiliary XSLT functions (see Appendix~\ref{appxslt}):
\begin{itemize}
\item
\texttt{pair}, taking a string parameter $s$,
and transforming an input document $d$
into the document $$ \Tree [.pair [.name $s$ ] $d$ ] $$
\item
\texttt{rewrite}, taking an XML parameter $p$ of the form
$$ \Tree [.cmb
[.pair [.name $s_1$ ] $d_1$ ] {\dots} [.pair [.name $s_n$ ] $d_n$ ]] $$
where $s_1$, \dots, $s_n$ are different strings and $d_1$, \dots, $d_n$ are
arbitrary XML documents,
and
transforming an input query $q$ by replacing every occurrence of an element
of the form $$ \Tree [.table $s_i$ ] $$
by a copy of $d_i$.
\end{itemize}
We then write:
\begin{verbatim}
select
  rewrite(l.Q, (select
                  CMB(pair(def,name))
                from Views))
from Log l
\end{verbatim}

\section{Semantical meta-querying} \label{semquery}

The language we have so far: SQL combined with XSLT, and enriched with XML
variables and XML aggregation, gives us all the power we need for ad-hoc
\emph{syntactical} meta-querying.  We now complete Meta-SQL so as to allow
\emph{semantical} meta-querying as well.

To this end, we add a function EVAL for dynamic evaluation of SQL
queries.  EVAL takes an SQL query (more correctly, its syntax tree in
XML format) as input, and returns the table resulting from executing
the query.  A call to EVAL can appear in an SQL expression anywhere a
table reference can; the resulting table can thus be ranged over by a
standard range variable.

As an example, suppose we are given a table \texttt{Customer} with two
attributes: \texttt{custid} of type string, and \texttt{query} of type XML.
The table holds queries asked by customers to the catalogue of a store.
Every query returns a table with attributes \texttt{item}, \texttt{price}, and
possibly others.
The following meta-query shows for every customer the maximum price of items
he requested:
\begin{verbatim}
select custid, max(t.price)
from Customer c, EVAL(c.query) t
group by custid
\end{verbatim}

EVAL is all we need, provided we have enough
information about the output scheme of the
stored queries we are evaluating.  For example, in the previous meta-query, we
are only interested in the \texttt{price} attribute, and we know that
every stored query evaluates to a table that indeed has a \texttt{price}
column.  But what if we are given an arbitrary collection
of stored queries without information about their output schemes?
They could even all have different output schemes!

Such a situation neatly fits the genre known as ``semistructured
data'' \cite{abs_book}:
data that has a structure (scheme), but we do not know it, and it can be
irregular.  Since XML is the standard format for semistructured data, and
since we already have XML variables in Meta-SQL, we can easily solve the
problem with an untyped version of EVAL\@.  This UEVAL function works just
like EVAL, except that the table resulting from the dynamic evaluation
of the query is presented as a set of XML documents.

More concretely, suppose a particular stored query evaluates to a table with
attributes $(A,B,C)$.  Then every output row $(a,b,c)$ is represented
as the XML document $$ \Tree [.row [.$A$ $a$ ] [.$B$ $b$ ] [.$C$ $c$ ]] $$
The resulting set of XML documents is
ranged over by an XML variable rather than a standard range variable;
this is a new use we make of XML variables in the Meta-SQL language.

As a simple example,
recall the second example meta-query from the Introduction:
``which queries in the log return an empty answer?''  To express this
meta-query, we write:
\begin{verbatim}
select Q from Log l
where not exists
  (select x from x in UEVAL(l.Q))
\end{verbatim}
Note that \texttt{x} is an XML variable, whereas \texttt{t} in the previous
example is a standard SQL range variable.

The fourth example meta-query from the Introduction (query comparison after
view expansion) is equally easy to express, given that we already saw how to
express view expansion in the previous section.  In particular, note
that we can apply EVAL and UEVAL not only to queries directly stored in the
database, but also to queries coming from a syntactical meta-subquery (such as
view expansion).

\paragraph{Run-time errors}  Both EVAL and UEVAL expect their input to be
(the XML syntax tree of) a correct SQL query.
If that is not the case, we consider this
to be a programming error, and a run-time exception
will be raised.  Moreover
an application of EVAL can still fail even if its argument is a correct
SQL query, namely,  when the
query result does not have the expected columns.

Designing a meta-query language where dynamic evaluation of stored queries
is statically typed so as to be safe from such run-time errors is possible
\cite{nvvv_typed},
but leads to an overly restricted formalism.  In the design
of Meta-SQL, we have opted to prefer flexibility and expressive power above
static typing.

\section{Implementation}
\label{sec:impl}
Meta-SQL is entirely compatible with modern
SQL implementations offered by contemporary extensible database systems.
Extensible (also called object-relational, or universal)
database systems \cite{stonebraker_wave}
support user-defined data types for the columns of tables,
and allow user-defined functions on these types to be called within SQL\@.
Extensibility is now part of the new SQL:1999 standard, and
the major commercial vendors are aggressively moving to support it.

\begin{figure*}
\resizebox{\textwidth}{!}{\includegraphics{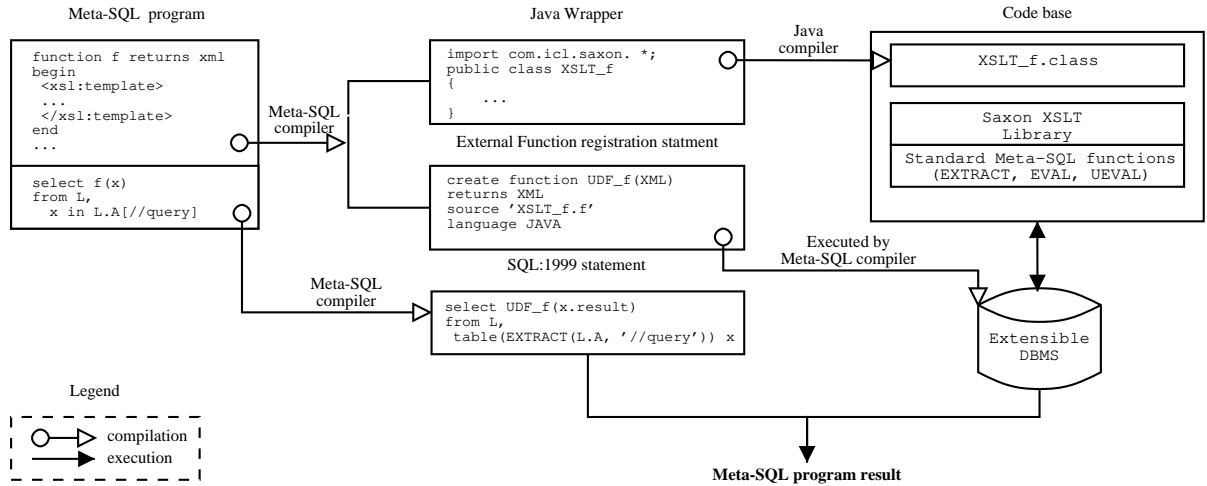}}
\caption{Implementing Meta-SQL.}
\label{architecture}
\end{figure*}

In Figure~\ref{architecture} we illustrate the architecture of an
implementation of Meta-SQL, explained in more detail in this section.

\paragraph{Implementing XML columns}
To support XML columns,
it suffices to define a data type `XML'.  We could derive
this type from the built-in type CLOB (Character Large Object) and store XML
documents as texts, but we could also derive from BLOB (Binary Large Object)
and store XML documents as binary encodings of their DOM tree structures.

\paragraph{Implementing XSLT calls}  Starting from
Meta-SQL source code, consisting of
a number of XSLT functions, followed by an SQL query using these functions, the
Meta-SQL compiler does the following automatically:
\begin{enumerate}
\item
For each XSLT function, we generate a wrapper function
(in an external programming language such as C or Java) that invokes an XSLT
processor, thus performing the required XSLT transformation on the arguments,
and returns the result.
If necessary, this result is converted to an SQL datatype like number or
string.
\item
All wrapper functions are compiled and put together in an object library.
\item
Every wrapper function is registered in the database system (using SQL:1999
CREATE FUNCTION statements).
\item \label{sqlstep}
With the functions in place,
the final Meta-SQL query can now be executed literally by the database system.
\end{enumerate}
Almost all available XSLT processors provide an interface to various
programming languages, and can be loaded together with the application that
invokes them.  Hence, the compiled wrapper functions can be loaded into the
database server together with the XSLT processor functionality and executed as
efficiently as possible.

\paragraph{Implementing XML variables}  Now the first and the last step in the
above plan become more involved:  in the first step, we must generate an
additional function \texttt{EXTRACT};  in the last step, we must
perform some rewriting on the SQL query.  We next explain this in a bit more
detail.

To support the XSLT calls, we needed only \emph{single-valued}
external functions: they return as output only a single value, be it a string,
a number, or an XML document.  To support XML variables, however,
we need a \emph{set-valued} external function.  Specifically, the Meta-SQL
system provides a function \texttt{EXTRACT},
which takes an XML document $y$ and an XPath expression $e$ as input,
and which returns the set of all subelements of $y$ that satisfy $e$.
This set is returned as a one-column table with attribute name
\texttt{result}.

Every XML variable binding, say, \texttt{$x$ in $y$[$e$]},
is now rewritten by the Meta-SQL compiler into
a call \texttt{EXTRACT($y$,'$e$')}.  This call returns a table to which
$x$ is again bound, but now $x$ has become
a standard SQL range variable over the single attribute \texttt{result}.

For example, recall the first example of Section~\ref{xmlvaragg}:
\begin{verbatim}
select v.name, string_value(x)
from Views v, x in v.def[//table]
\end{verbatim}
This query will be rewritten as follows:
\begin{verbatim}
select v.name, string_value(x.result)
from Views v,
     table(EXTRACT(v.def,'//table')) x
\end{verbatim}
The above from-clause contains two table expressions.
Note that the variable ranging over the first table expression, namely
\texttt{v}, is directly used in the second table expression.  This is an
instance of what SQL:1999 calls a ``lateral derived table''
\cite{melton_sql99}.
Such lateral joins were not allowed in SQL-92; we see that they are crucial
here.  We point out that they were present in OQL
from the outset \cite{cluet_oql}.

\paragraph{Implementing EVAL and UEVAL}
The function UEVAL can, like EXTRACT, be realized as a set-valued external
function.
This evaluation function takes an SQL syntax tree in XML as input;
unparses it; sends the SQL query to the database; receives the answer
rows; transforms them into XML as explained in the previous section; and
returns the results.  The XML variable ranging over the UEVAL result is
handled in the same way as above.

For example,
\begin{verbatim}
select Q from Log l
where not exists
  (select x from x in UEVAL(l.Q))
\end{verbatim}
is compiled into
\begin{verbatim}
select Q from Log l
where not exists
  (select x.result
   from table(UEVAL(l.Q)) x)
\end{verbatim}

The implementation of EVAL is a bit more complicated, because EVAL returns not
a set of XML documents, but a normal SQL table ranged over by a standard range
variable.  In this case, the Meta-SQL compiler first determines
the specific attributes that are mentioned
in connection with this range variable.
A specific table-valued external function having this set of attributes as
output scheme is then generated and registered in the database system.
This evaluation function will actually not send its exact
argument SQL query
to the database for evaluation, but rather its projection on the output
scheme in question.

For example, recall the first example of Section~\ref{semquery}:
\begin{verbatim}
select custid, max(t.price)
from Customer c, EVAL(c.query) t
group by custid
\end{verbatim}
To implement this query, the Meta-SQL system will generate an evaluation
function, say, EVAL\_1,
with output scheme \texttt{(price)}, and will rewrite the
query to
\begin{verbatim}
select custid, max(t.price)
from Customer c,
     table(EVAL_1(c.query)) t
group by custid
\end{verbatim}
Here,
${\rm EVAL\_1}(q)$,
when called on any stored query $q$,
evaluates and returns the projection
$\pi_{\text{\texttt{price}}}(q)$.

If the select-clause would have additionally included \texttt{sum(t.qty)},
then the system would have generated a different evaluation function
EVAL\_2 with output scheme \texttt{(price,qty)} and according behavior.

\paragraph{Implementing XML aggregation}  The XML aggregate function CMB
introduced in Section~\ref{xmlvaragg} can be directly provided as a
user-defined aggregate function.

\paragraph{A working prototype} We have developed
a prototype implementation on top of DB2 UDB
\cite{chamberlin_udb}, which we chose because it is freely available
to university research and teaching.  

We implement the external XSLT wrapper functions in Java, in
conjunction with the popular and free Java-based XSLT processor SAXON
\cite{saxon}.  SAXON also provides a convenient Java-XML interface
making the EXTRACT function very easy to write.\footnote{IBM provides
  an `XML Extender' to DB2 UDB which already provides an XML data type
  (derived from CLOB as we do), but with insufficient functionality
  for our needs.  For example, there is also an Extract function, but
  it is much weaker than the EXTRACT function we need to implement our
  mechanism of XML variables.}  A caveat in connection to the
implementation of EVAL and UEVAL is that DB2 UDB currently does not
allow SQL calls inside external functions.  We bypass this restriction
by spawning a child process, which then makes a new connection to the
database in order to evaluate the stored query.

\section{Experimental performance evaluation}
\label{sec:experiments}
In this section we describe some performance experiments on our
prototype implementation. Unless stated otherwise, the results shown
are averages of multiple executions of the test under discussion.

\paragraph{Java Overhead}
Our first test measures the overhead implied by calling external
functions written in Java. To that cause we created three simple
single-valued external functions, \texttt{CST}, \texttt{MUL} and
\texttt{CAP}, which respectively return a constant number, multiply a
number with a constant, and transform the input string to capital
letters. The following four queries were executed:
\begin{verbatim}
select * from T
select CST(A), B from T
select MUL(A), B from T
select A, CAP(B) from T
\end{verbatim}
Here, \verb!T! consists of an integer column \texttt{A} and a
character column \texttt{B}. The first query is executed to measure
the time needed to select tuples from \texttt{T} without calling an
external function.

\begin{figure*}[tbp]
  \centering
  \scalebox{0.75}{\resizebox{\textwidth}{!}{\includegraphics{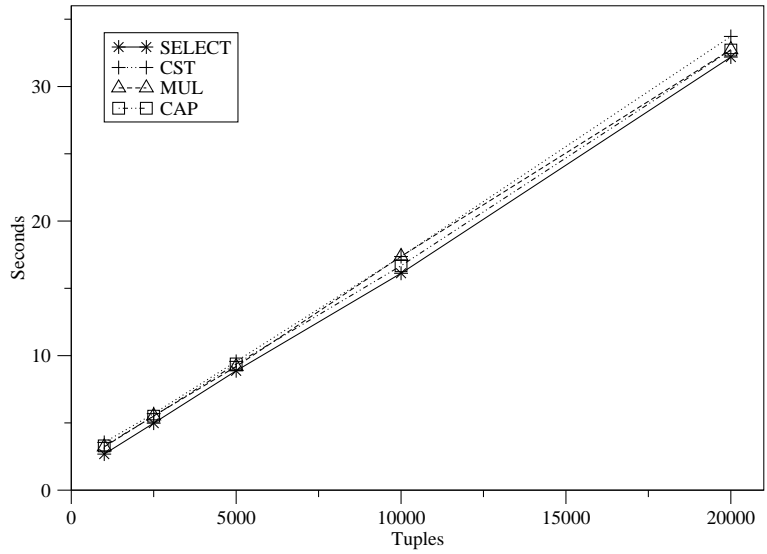}}}
  \caption{Performance of external functions in Java}
  \label{fig:javac}
\end{figure*}

Figure \ref{fig:javac} shows the results for varying table sizes. As
was to be expected, the running times of all functions grow linearly
in the number of input tuples. Since the time needed to execute the
last three queries closely resembles that of the first, we may
conclude that there is a minimal overhead involved with external
functions programmed in Java.

\paragraph{XSLT Processor overhead}

Our prototype implements XSLT functions by external wrapper functions
calling the SAXON XSLT processor. To measure the startup cost of the
XSLT processor, we compared the Meta-SQL query
\begin{verbatim}
function XSLT_CST returns number
begin
  <xsl:template match="/">
    40513
  </xsl:template>
end

select XSLT_CST(A) from T
\end{verbatim}
whose execution time is dominated by starting up the XSLT processor,
with the query
\begin{verbatim}
select CST(A) from T
\end{verbatim}
Here, \verb!CST! is  single-valued external function which also
returns the same constant on all inputs. Table \verb!T! consists of a
single XML column, whose values are single-node trees. 

\begin{figure*}[tbp]
  \centering
  \scalebox{0.75}{\resizebox{\textwidth}{!}{\includegraphics{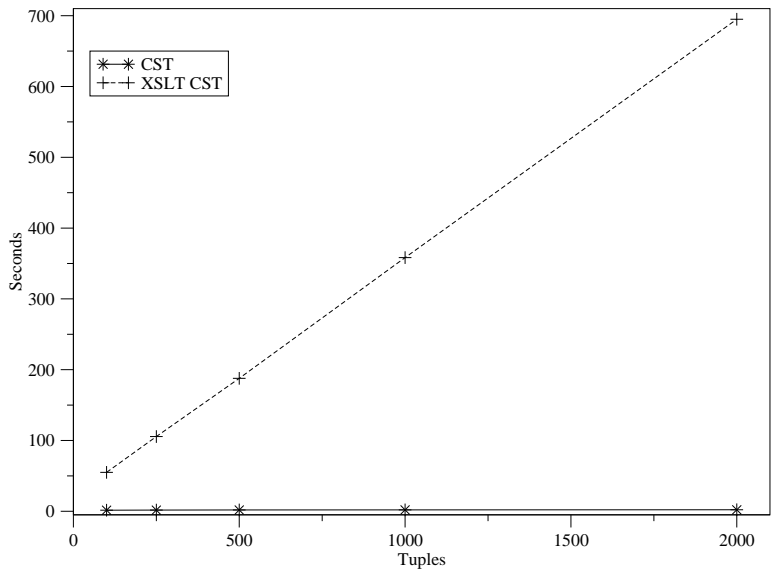}}}
  \caption{Performance of XSLT calls}
  \label{fig:xslt-test1}
\end{figure*}

Figure \ref{fig:xslt-test1} shows the results for varying table sizes.
The XSLT function running time is significantly larger than that of
\texttt{CST}, and grows linearly in the number of input tuples, which
indicates that there is a constant startup cost imposed by the XSLT
processor. As such, a Meta-SQL implementation would greatly benefit
from more efficient XSLT/XML processors.

A valid question to ask next is how the size of the input XML document
affects the running time of an XSLT function.  To this cause we
compared the following two queries:
\begin{verbatim}
select XSLT_COPY(A) from T1
select COPY(A) from T1
\end{verbatim}
Here, both \verb!XSLT_COPY! and \verb!COPY! copy their input to the
output. The only difference is that XSLT function \verb!XSLT_COPY!
does so by XSLT template matching whereas external function
\verb!COPY! performs a true copy.  Table \verb!T1! consists of a
single XML column \verb!A!.

\begin{figure*}[tbp]
  \centering
  \scalebox{0.75}{\resizebox{\textwidth}{!}{\includegraphics{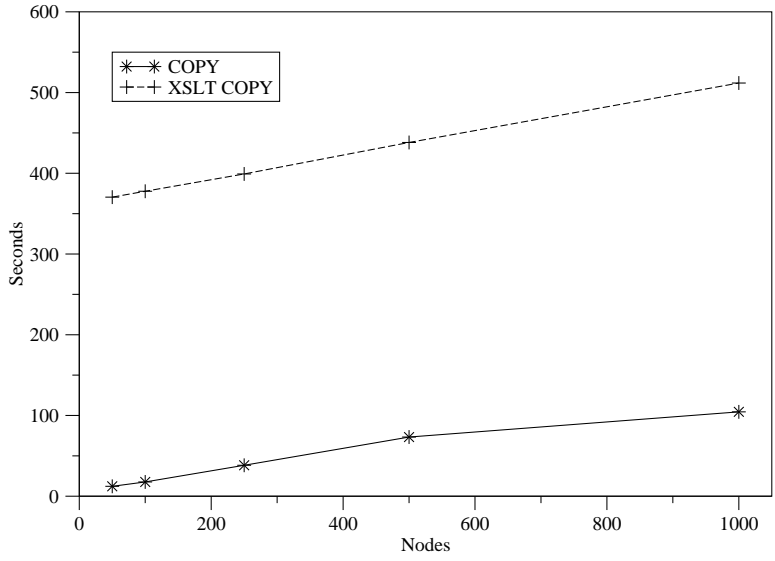}}}
  \caption{Performance of XSLT calls}
  \label{fig:xslt-test2}
\end{figure*}

Figure \ref{fig:xslt-test2} shows the running times of these queries
on tables with $1000$ documents, for varying document sizes. Both
running times grow linearly when the number of nodes per tuple
increases, as was to be expected. Combined with the previous results,
this indicates that calling a XSLT function has a rather large
start-up cost, but a relatively small execution cost.

\paragraph{Extract}
As described in the Section \ref{sec:impl}, XML variables are implemented by
a set-valued external function called \texttt{EXTRACT}\@. To see how
set-valued external functions compare with single valued functions, we
created a single-valued function named \verb!SCALAR_EXTRACT!, which
evaluates the XPath expression \verb!/! on its input, and executed the
following three queries:
\begin{verbatim}
select x from T, x in B[//*]
select x from T1, x in B[/] 
select SCALAR_EXTRACT(B) from T1
\end{verbatim}
Here, \verb!T! consists of an XML column \verb!B!, which is populated
by $1000$ documents of $m$ nodes. Table \verb!T1! contains the result
of the first query. Hence, the second and third query return the same
output as the first; here, the second query still uses an XML
variable, whereas the third query makes a direct function call. The
difference is that \verb!EXTRACT!  is called $1000$ times in the first
query, each time returning $m$ tuples, whereas in the second and third
query \texttt{EXTRACT} and \texttt{SCALAR\_EXTRACT} are called $m
\times 1000$ times, each call returning a single tuple. Thus, this
test compares the overheads involved with set-valued functions
returning multiple tuples, set-valued functions returning a single
tuple, and single-valued functions returning a single tuple.

\begin{figure*}[tbp]
  \centering
  \scalebox{0.75}{\resizebox{\textwidth}{!}{\includegraphics{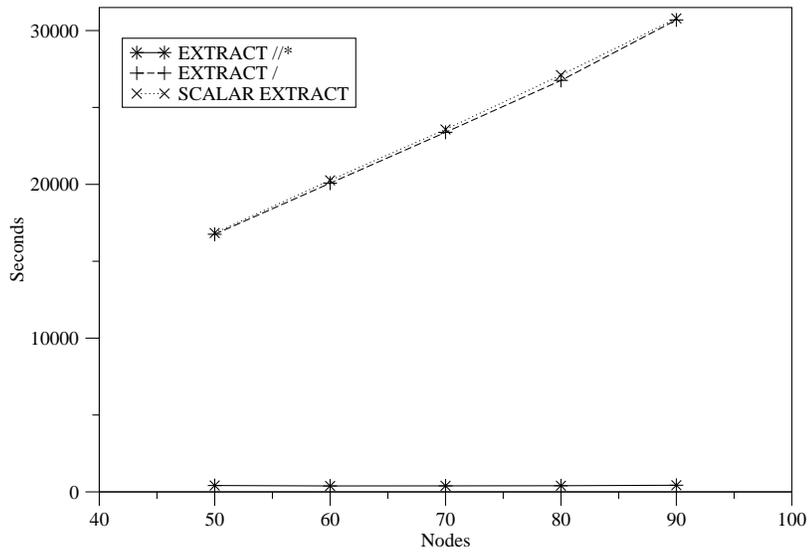}}}
  
  \caption{Performance of Extract---Experiment 1}
  \label{fig:extract-test1}
\end{figure*}

We timed the execution for varying $m$, as shown in Figure
\ref{fig:extract-test1}. Due to the large running times, the test was
only timed once.

We can safely conclude that returning multiple tuples from a
set-valued function is not a problem. Indeed, the running time of the
first query is quite reasonable and increases linearly with $m$
(although it appears constant in Figure \ref{fig:extract-test1} due to
the large running times of the other queries). In particular, it is
multiple times faster than returning $m$ times a single tuple. As
such, the implementation of XML variables in our approach performs
very reasonably.

Since both \texttt{EXTRACT} and \texttt{SCALAR\_EXTRACT} use the XSLT
processor to evaluate their XPATH expressions and because the running
time of the second query resembles that of the third, we can also
conclude that the overhead of a set-valued external function is about
the same as that of a single-valued one.

With the following two queries, we measure the overhead of an
\texttt{EXTRACT} call versus the actual amount of work that needs to
be done:
\begin{verbatim}
select x from T, x in B[//*]
select C from T, T2
\end{verbatim}
Here, \verb!T! is as before and \verb!T2! is the table with XML column
\verb!C!, containing the result of the first query on a single
document in \texttt{T}.  Since all documents in \texttt{T} are the
same, both queries return the same result. By adding the time the XSLT
processor needs to evaluate the XPath expression \verb!//*! $1000$
times to the timing of the second query, we get an estimate of the
time needed to calculate the result of the first query, without
actually calling \texttt{EXTRACT}.

\begin{figure*}[tbp]
   \centering
  \scalebox{0.75}{\resizebox{\textwidth}{!}{\includegraphics{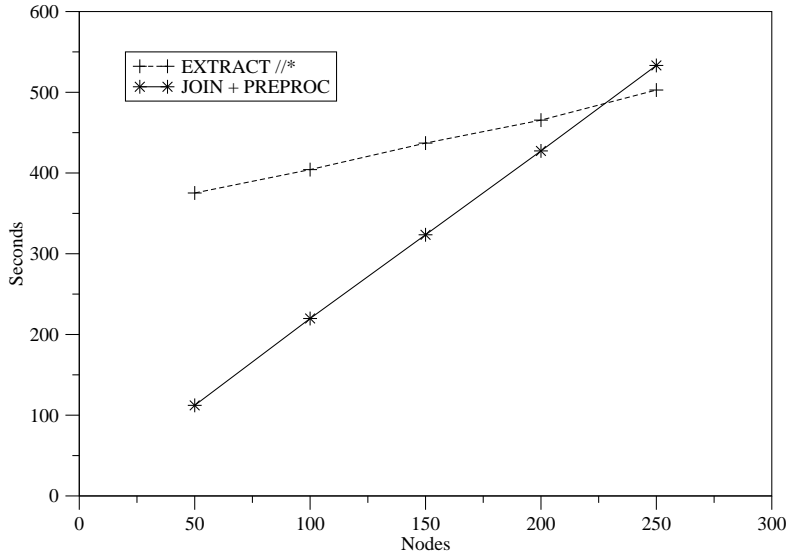}}}

   \caption{Performance of Extract---Experiment 2}
   \label{fig:extract-test2}
 \end{figure*}

As can be seen from Figure \ref{fig:extract-test2}, the running time
of both queries grow linearly in their input. However, the first query
outperforms the second one when $m$ grows larger. Hence, although
\texttt{EXTRACT} has some startup overhead, it is efficient when
applied to large documents. Indeed, it even outperforms a setting in
which the same amount of work needs to be done, but no call to a
set-valued external function is made. Combined with our previous
results, this shows that XML variables are efficiently implementable.

\paragraph{Eval}
Remember from Section \ref{sec:impl} that \texttt{EVAL} and
\texttt{UEVAL} are implemented as set-valued external functions which
create a subprocess to execute their input query. In order to measure
the overhead incurred by starting this process and communicating with
it, we created tables \texttt{T} and \texttt{T'} with integer column
$A$ and XML column $B$.  Every tuple in \texttt{T} contains in its $B$
column the following query in XML format:
\begin{verbatim}
select * from T1
\end{verbatim}
We then measured the time needed to execute the query
\begin{verbatim}
select A, e.A
from T, EVAL(T.B) as e
\end{verbatim}
Furthermore, with $n$ be the number of tuples in $T$, we timed the
execution of $n$ times the query 
\begin{verbatim}
select * from T1
\end{verbatim}
(in Java, since \texttt{EVAL} uses Java to execute its queries) and added that
to the time needed to compute the same result:
\begin{verbatim}
select T.A, T1.A from T, T1
\end{verbatim}

Comparison of these two measurements gives a good indication of the
overhead implied by \texttt{EVAL}. 

\begin{figure*}[tbp]
   \centering
   \scalebox{0.75}{\resizebox{\textwidth}{!}{\includegraphics{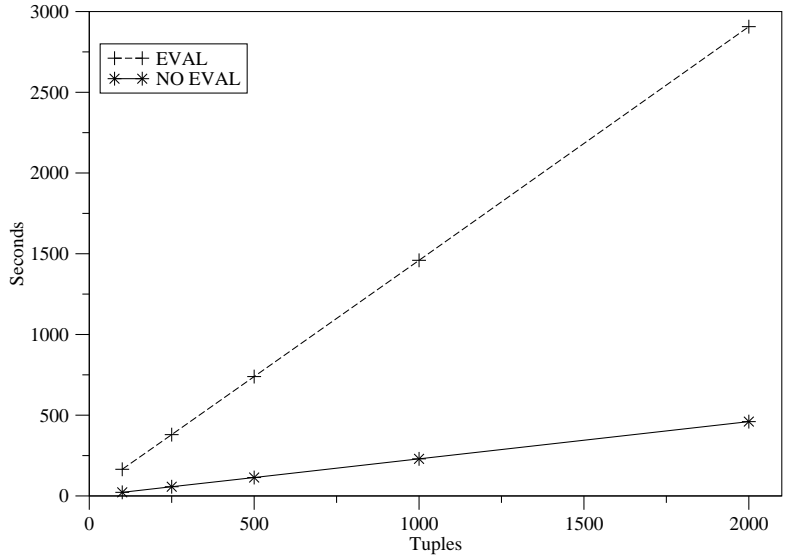}}}

   \caption{Performance of Eval}
   \label{plot:eval}
 \end{figure*}

As shown in Figure \ref{plot:eval}, calling the \texttt{EVAL} function
is six to seven times slower than preprocessing and joining, and grows
linearly in the number of input tuples.

\paragraph{Conclusion}
The experiments above indicate that our proposal for implementing
meta-querying features \emph{on top of} the database engine induces
constant, predictable overheads, which is certainly good news. 

Still, the running times are sometimes too high. Therefore, more
efficient XSLT processors would certainly help. The ideal solution
would be to incorporate XSLT processing directly into the query
processor, as has recently been suggested by Moerkotte
\cite{moerkotte_xslt}. It would also help if we could implement the
EVAL and UEVAL functions entirely in a programmed SQL language such as
Oracle's PL/SQL, or SQL/PSM, in which case the overhead of calling a
table function and starting a subprocess could be avoided.

On the other hand, our current implementation method has the advantage
that it is generally applicable without a need to change the internal
query processor (which is often not feasible).

\section{Discussion}

\textbf{Other meta-query languages.}  Meta-SQL is the first practical
language for meta-querying.  At the same time, however, it is firmly
based on our past experience in designing formalisms for
meta-querying.  More specifically, two of us (in collaboration with
Neven and Van Gucht) have earlier introduced the \emph{reflective}
relational algebra \cite{vvv_reflect} and the \emph{meta} relational
algebra \cite{nvvv_typed}, two formal meta-query languages based on
the relational algebra rather than on SQL\@.  The two formalisms
differ in their approach: the reflective algebra is untyped, stores
queries in so-called ``program relations'', and uses the basic
relational algebra operators for the syntactical manipulation; in
contrast, the meta algebra is statically typed and views stored
queries as an abstract data type with specific operations for
syntactical manipulation.  Meta-SQL is the practical culmination of
these two proposals, and stands in between the two approaches: like
the meta algebra, it stores queries in an XML abstract data type, but
like the reflective algebra, it is essentially untyped.  Every
meta-query expressed in the reflective relational algebra, or in the
meta relational algebra, can also be expressed in Meta-SQL.

\textbf{Schema querying.}
Starting with the seminal paper on HiLog \cite{ckw_hilog}, the concept of
\emph{schema querying} has received considerable attention in the
literature.
Clearly, schema querying is a special kind of meta-querying.
For instance, SchemaSQL \cite{schemasql} augments SQL with
generic variables ranging over table names, rows, and column names.
It is not difficult to simulate SchemaSQL in Meta-SQL, using XML variables and
UEVAL calls.  Of course, since SchemaSQL is more specialized, it allows more
opportunities for optimized implementation \cite{schemasql2}.

\textbf{Commercial XML extensions.} All the major commercial
ORDBMS vendors are providing XML extensions to their products,
However, the emphasis there is mainly on publishing results of SQL
queries in XML format (e.g.,
\cite{almaden_relationalxml,silkroute}), so that they can be
further processed using the standard XML tools, including XSLT\@.
This combination of SQL and XSLT is clearly quite different in
scope compared to the combination we have proposed in this paper.
The other direction, where large XML documents are decomposed and
stored in tabular format, has also been researched (e.g.,
\cite{niagara_xmlrelational,stored}) and is getting into the
commercial products.

We also mention that several features of ``SQL/XML'' \cite{sqlx,sqlx2}
such as the XML data type and the mapping of tables to XML values are
similar to those found in this paper.  Moreover, an Extract-like
operator and functions operating on XML values (albeit through XQuery)
are listed among the future work.

\textbf{Text extensions.}  Given that most database systems now support
a text data type with better functions for text searching and editing than
standard SQL, one may wonder why we cannot support meta-querying simply by
representing the stored queries as text.  The answer is that for many
meta-queries the \emph{structure} of the stored
queries is important.  For example, searching for the use of a certain view
name in a query is more than a simple pattern search.  Using syntax
trees and XSLT makes structural querying very easy.

\textbf{XML query languages.} Although the focus of this paper has been on
meta-querying as opposed to general XML querying, we still would like
to conclude by pointing out that Meta-SQL, without EVAL, is
not only a language for syntactical meta-querying, but can serve in general
as a query language for databases containing XML documents in addition to
ordinary relational data.  Its closeness to standard SQL and
object-relational processing is a major advantage.
On the other hand,
the treatment of XML documents as abstract data items
seems less appropriate for ``pure'' XML
databases, i.e., huge XML documents.  However, as already indicated above,
there seems to be a strong trend in the database processing world to decompose
such XML databases into relational data anyway.

\paragraph{Acknowledgment}
We are indebted to Frank Neven and Dirk Van Gucht for inspiring conversations.
We also thank our students Igor Kalders, Frank Pilgrim, and Jef Vos for their
help with the prototype implementation.

\appendix

\section{A DTD for SQL} \label{appdtd}

The following is a reasonably complete DTD for syntax trees of SQL-92
select-statements, with the exception of the various join operators.
The DTD is derived from the grammar given by Date \cite{date_sql92}.

\begin{verbatim}
<!ELEMENT query
 ((select, from, where?, group_by?,
   having?)
 | (union | except | intersect))>
<!ELEMENT select
 ((all | distinct)?,
  (wildcard | sel-item+))>
<!ELEMENT all EMPTY>
<!ELEMENT distinct EMPTY>
<!ELEMENT wildcard EMPTY>
<!ELEMENT sel-item
 ((column
  | (rangevar, (column | wildcard)))
  | scalar
  | aggregate)
 , alias?)>
<!ELEMENT rangevar (#PCDATA)>
<!ELEMENT column (#PCDATA)>
<!ELEMENT column-ref
 (rangevar?, column)>
<!ELEMENT scalar
 (alg-exp | concat-exp | column-ref
 | aggregate | constant | query)>
<!ELEMENT aggregate
 (count-all
 | ((avg | count | max | min | sum),
    (all | distinct)?,
    (alg-exp | concat-exp
    | column-ref | constant
    | query)))>
<!ELEMENT count-all EMPTY>
<!ELEMENT avg EMPTY>
<!ELEMENT count EMPTY>
<!ELEMENT max EMPTY>
<!ELEMENT min EMPTY>
<!ELEMENT sum EMPTY>
<!ELEMENT alg-exp
 (scalar, (add | sub | mul | div),
  scalar)>
<!ELEMENT add EMPTY>
<!ELEMENT sub EMPTY>
<!ELEMENT mul EMPTY>
<!ELEMENT div EMPTY>
<!ELEMENT concat-exp
 (scalar, scalar)>
<!ELEMENT constant (#PCDATA)>
<!ELEMENT from (table-ref+)>
<!ELEMENT table-ref
 ((table | query), alias?)>
<!ELEMENT alias (#PCDATA)>
<!ELEMENT table (#PCDATA)>
<!ELEMENT where (cond-exp)>
<!ELEMENT cond-exp
 (not?, (cond-test | and | or))>
<!ELEMENT not EMPTY>
<!ELEMENT cond-test
 (comparison | like | in
 | match | all-or-any | exists
 | unique | overlaps
 | test-for-null)>
<!ELEMENT and (cond-exp, cond-exp+)>
<!ELEMENT or (cond-exp, cond-exp+)>
<!ELEMENT rowconstr
 (column-ref | scalar)+>
<!ELEMENT comparison
 (rowconstr,
  (eq | lt | let | gt | get | neq),
  rowconstr)>
<!ELEMENT eq EMPTY>
<!ELEMENT lt EMPTY>
<!ELEMENT let EMPTY>
<!ELEMENT gt EMPTY>
<!ELEMENT get EMPTY>
<!ELEMENT neq EMPTY>
<!ELEMENT like
 ((column-ref | scalar),
  (column-ref | scalar),
  (column-ref | scalar)?)>
<!ELEMENT in
 ((rowconstr, query)
 | (scalar, scalar+))>
<!ELEMENT partial EMPTY>
<!ELEMENT full EMPTY>
<!ELEMENT match 
 (rowconstr, unique?, 
  (partial|full)?, query)>
<!ELEMENT all-or-any
 (rowconstr,
  (eq | lt | let | gt | get | neq),
  (all | any)?,
  query)>
<!ELEMENT any EMPTY>
<!ELEMENT exists (query)>
<!ELEMENT unique (query)>
<!ELEMENT overlaps
 (scalar, scalar, scalar, scalar)>
<!ELEMENT test-for-null (rowconstr)>
<!ELEMENT group-by (column-ref+)>
<!ELEMENT having (cond-exp)>
<!ELEMENT union (query, all?, query)>
<!ELEMENT except
 (query, all?, query)>
<!ELEMENT intersect
 (query, all?, query)>
\end{verbatim}

\section{XSLT programs} \label{appxslt}
\begin{verbatim}
function cartprod returns xml
begin
<xsl:template match="/">
 <query>
  <select> <wildcard/> </select>
  <from>
   <xsl:apply-templates
          select="cmb/*"/>
  </from>
 </query>
</xsl:template>
<xsl:template match="/cmb/*">
 <table-ref>
   <xsl:copy-of select="."/>
 </table-ref>
</xsl:template>
end

function pair
param s string
returns xml
begin
<xsl:param name="s"/>
<xsl:template match="/">
 <pair>
  <name>
   <xsl:value-of select="$s"/>
  </name>
  <xsl:copy-of select="*"/>
 </pair>
</xsl:template>
end

function rewrite
param p xml
returns xml
begin
<xsl:param name="p"/>
<xsl:template match="*">
 <xsl:copy>
  <xsl:apply-templates/>
 </xsl:copy>
</xsl:template>
<xsl:template match="table">
 <xsl:apply-templates select="$p"
                 mode="find">
  <xsl:with-param name="search"
                 select="string(.)"/>
  <xsl:with-param name="caller">
   <xsl:copy-of select="."/>
  </xsl:with-param>
 </xsl:apply-templates>
</xsl:template>
<xsl:template match="/" mode="find">
 <xsl:param name="search"/>
 <xsl:param name="caller"/>
 <xsl:param name="found"
  select="cmb/pair[name=$search]"/>
 <xsl:choose>
  <xsl:when test="$found">
   <xsl:copy-of
     select="$found/query"/>
  </xsl:when>
  <xsl:otherwise>
   <xsl:copy-of select="$caller"/>
  </xsl:otherwise>
 </xsl:choose>
</xsl:template>
end
\end{verbatim}

\end{document}